\documentclass[structabstract]{aa} 
\usepackage{graphicx}
\usepackage{aalongtable}
\usepackage{lscape}
\usepackage{txfonts}
\begin{document}
\title{Long-period variables in \object{NGC\,147} and \object{NGC\,185}
\thanks{Table \ref{periods147} and \ref{periods185} are only available in electronic 
form via http://www.edpscience.org}}
\author{D. Lorenz\inst{1}
\and T. Lebzelter\inst{1}
\and W. Nowotny\inst{1}
\and J. Telting\inst{2} 
\and F. Kerschbaum\inst{1} 
\and H. Olofsson\inst{3,4} 
\and H.E. Schwarz\inst{5}
}
\institute{University of Vienna, Department of Astronomy, 
T\"urkenschanzstrasse 17, A-1180 Vienna, Austria\\
\email{denise.lorenz@univie.ac.at}
\and
Nordic Optical Telescope, Apartado 474, 
38700 Santa Cruz de La Palma, Spain\
\and
Department of Astronomy, Stockholm University, 
AlbaNova University Center, 10691 Stockholm, Sweden\
\and
Onsala Space Observatory, 43992 Onsala, Sweden\
\and
Deceased, 2006 October 20}
\date{Received ; accepted } 
\abstract
{Previous studies on the stellar content of the two nearby dwarf galaxies 
\object{NGC\,147} and \object{NGC\,185} reveal a rich population of late-type giants in both
systems, including a large number of carbon-rich objects.
These stars are known to show pronounced photometric variability, which can be 
used for a more detailed characterisation of these highly evolved stars. Owing to
their well-studied parameters, these Local Group 
members are ideal candidates for comparative studies.}
{Trough photometric monitoring, we attempt to provide a catalogue of 
long-period variables (LPVs), including Mira variables, semi-regular variables,
and even irregular variables in \object{NGC\,147} and \object{NGC\,185}. We investigate the
light variations and compare the characteristics of these two 
LPV populations with the results found for other galaxies such as the LMC.}
{We carried out time-series photometry in the $i$-band of the two target 
galaxies with the Nordic Optical Telescope (NOT), covering a time span of
$\approx$\,2.5\,years. More than 30 epochs were available for a period search. 
These data were then combined with single-epoch $K$-band photometry, also
obtained with the NOT. Narrow-band photometry data from the literature was used
to distinguish between O-rich and C-rich stars.}
{We report the detection of 513 LPVs in \object{NGC\,185} and 213 LPVs in \object{NGC\,147}, 
showing amplitudes $\Delta i$ of up to $\approx$\,2$^{\rm mag}$
and periods ranging between 90 and 800 days. The period-luminosity diagram 
for each of our target galaxies exhibits a well populated 
sequence of fundamental mode pulsators. The resulting period-luminosity
relations we obtained are compared to relations from the literature.
We discuss the universality of those relations because of which, as a side result,
a correction of the distance modulus of \object{NGC\,185} may be necessary. A value of 
($m$--$M$)=$24\fm30$ seems to be more appropriate to match the observed data.
Only one of our two galaxies, namely \object{NGC\,185}, has a significant fraction of 
possibly first overtone pulsators. An interpretation of this finding in terms of
differences in the star-formation histories is suggested.
}
{}
\maketitle
%
\section{Introduction}\label{s:intro}

Asymptotic Giant Branch (AGB) stars are highly evolved stars with low to 
intermediate initial masses of $\approx$\,$0.6-8 M_{\odot}$ that have passed the
helium-core burning phase. These stars are then powered by nuclear burning of
hydrogen and helium in two thin shells on top of a core of carbon (C) and oxygen
(O). During the early AGB phase these stars are O-rich, showing a photospheric
C/O-ratio\,$<$\,1, and most of them can be classified as stars of spectral type
M. For AGB stars with initial masses up to $\approx$4\,$M_\odot$ the atmospheric
chemical composition can change dramatically because processed elements, most
notably \element[][12]{C}, are dredged up to the surface by convective mixing
after a thermal pulse. Depending on the C/O-ratio, their spectral type changes
from K or M via S to C (C/O\,$\geq$\,1) (Groenewegen \cite{groene2}). For AGB
stars with initial masses $\gtrsim 4M_{\odot}$ the temperature at the bottom of
the convective envelope rises sufficiently high to transform C into N. This process,
called hot-bottom-burning, causes some AGB star to remain O-rich. 

\begin{figure*}
\resizebox{\hsize}{!}{\includegraphics{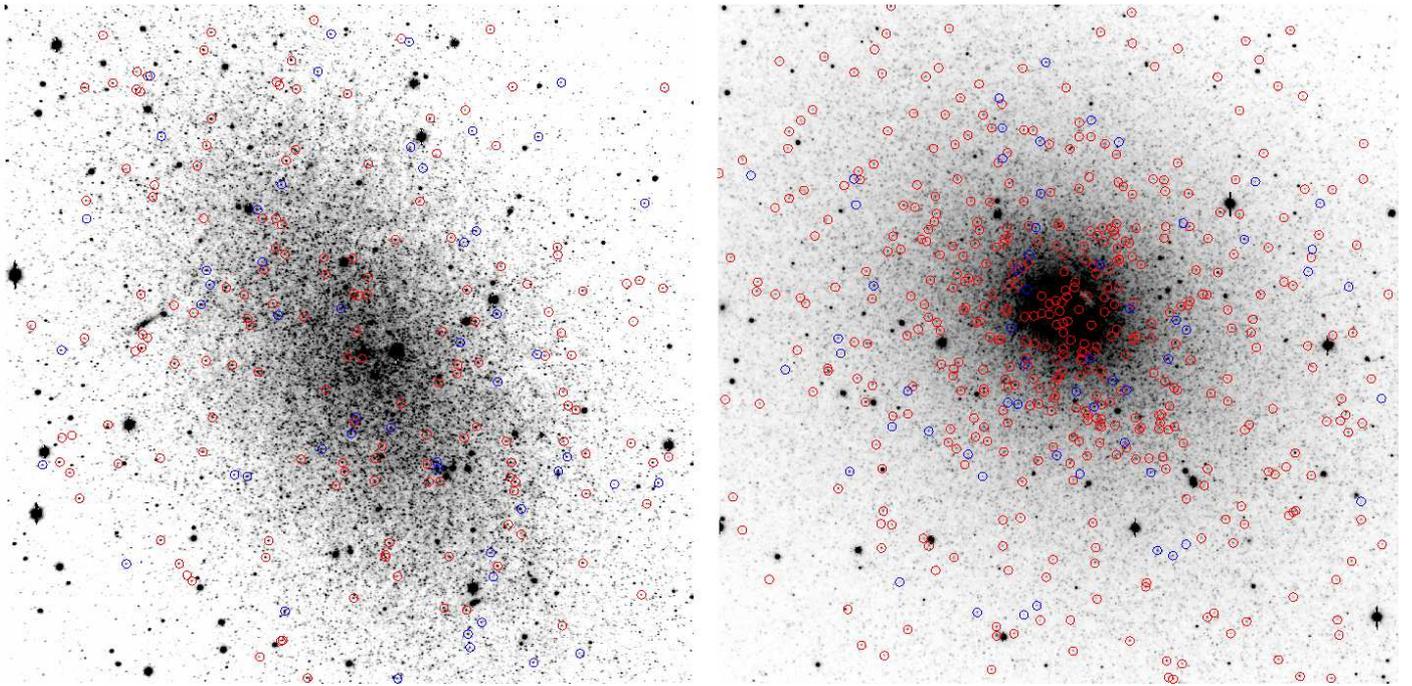}}
\caption{
Inverted CCD images of NGC147 (left panel) and NGC185 (right panel) obtained with 
NOT in the i-band. North is up and east is to the left. Designated with blue 
circles are objects that were identified as LPVs in this work and for which the  
narrow-band photometry in Paper\,I indicates \mbox{C-rich} atmospheric chemistry. 
The red circles mark all other stars (presumably \mbox{O-rich}) that were found to show 
long-period variability.
}
\label{GalLPV}
\end{figure*} 

Long-period variables (LPVs) is the generic term for variable stars known as
Mira variables and semi-regular variables. They generally show periodic 
variations in brightness with
periods of $\approx$30 up to a few thousands of days and amplitudes ranging from 
several tenths to approximately ten magnitudes in the visual. 
By studying LPVs in the
Large Magellanic Cloud (LMC), Wood et al. (\cite{woodb}) and Wood (\cite{woodc})
were the first to discover that all LPVs seem to group around at least five
almost parallel sequences in a period-luminosity-diagram (PLD), to which the authors 
assigned the letters A to E. Sequence\,C was explained consisting of stars
that pulsate in fundamental mode. Long-period variables on sequence\,B were 
explained as first and
second overtone pulsators, and stars on sequence\,A as higher overtone
variables. Later publications of period-luminosity relations (PLRs) of the 
LMC revealed that sequence\,B actually splits into two separate sequences, namely 
B and C$^{\prime}$ (see Kiss \& Bedding \cite{kiss}; Ita et al. \cite{ita}). 
Meanwhile, different authors (e.g. Fraser et al. \cite{fraser}; Soszy\'{n}ski et 
al. \cite{soszy_a}) are using different labels in a PLD; we use the original 
labelling of Wood et al. (\cite{woodb}) plus the additional sequence\,C$^{\prime}$. 
For stars belonging to sequences E or D the variability cannot be attributed to
radial pulsation. Stars on sequence\,E are thought to be related to close binary
systems showing ellipsoidal variability (see Soszy\'{n}sky et al.
\cite{soszy_b}). Recently, Nicholls et al. (\cite{nich}) confirmed this
assumption by comparing the phased light and radial velocity curves of LMC red
giant binaries. The results of this study also demonstrated that the variations
of stars on sequence\,E and D are caused by different mechanisms. The LPVs on
sequence D show periodicites on two time scales, where the long secondary period
(LSP) is about ten times the shorter period. The origin of these LSPs is still
unknown (Nicholls et al. \cite{nich2}). By comparing a sample of sequence\,D stars with
similar red giants (not showing LSPs), Wood \& Nicholls (\cite{wooni}) found
that such objects have a significant excess in the mid-infrared 
($8-24$\,$\mu$m), which is thought to originate from circumstellar dust. 
This is not the case for sequence\,E stars (Nicholls et al. \cite{nich}).
Within the last decade, several studies were carried out to explore PLRs of LPVs in different 
stellar systems of the Local Group (see Groenewegen \cite{groene}
for an overview). Rejkuba et al. (\cite{rejka}) and Rejkuba (\cite{rejk}) were
able to study LPVs even beyond the Local Group, namely in NGC\,5128 in the
Centaurus group. This growing sample of PLRs for LPVs raises the question
whether these relations are universal or not and if they are indeed universal, they can, 
therefore, be used as an additional tool to measure distances. Here we aim to contribute to this
discussion by investigating LPVs in the two dwarf galaxies \object{NGC\,147} and
\object{NGC\,185}. 

The two target galaxies of our investigation, \object{NGC\,147} and \object{NGC\,185},
were discovered by J. Herschel in September 1829 and by W. Herschel in November
1787, respectively, and are known to be members of the M\,31 subgroup. Together
with NGC\,205, they are the most luminous dSphs in the Local Group and are located at
an angular distance of approximately $12^\circ$ from the Andromeda nebula 
(van den Bergh \cite{vanbergh}; Corradi \cite{corr}). According to van den Bergh
(\cite{vanbergh}), they are separated by only 58\arcmin on the sky without any 
indication of interaction (Battinelli \& Demers \cite{batti_a}; Geha et al. 
\cite{geha}). Although these galaxies appear fairly similar concerning their
colour-magnitude diagrams (CMDs), some differences can be found in their star
formation histories (SFHs), most notably for recent epochs ($<$1\,Gyr see 
Mateo \cite{mateo}). Because there are no main-sequence turn-off stars with 
$M_{V}$\,$<$\,$-1$, Han et al. (\cite{han}) mentioned that the most recent large-scale 
star-forming activity in \object{NGC\,147} occurred at least 1\,Gyr in the past. 
According to broad-band near-infrared CMDs of Riebel (\cite{riebel}), this event 
happened $\approx$3\,Gyr ago and Dolphin (\cite{dolphin}) derived 4\,Gyr using HST
images. That there are no signs of dust and gas (Young and Lo
\cite{yolo}, Sage \cite{sage}) in this galaxy, which could serve as building
material for new stars, also supports the idea that star formation ceased long
ago. Using the relation between the $K$-magnitude of the AGB tip and age (as
predicted from isochrones; Girardi et al. \cite{gir}), Sohn et al. (\cite{sohn}) find that
most of the M-giants in \object{NGC\,147} formed between at $log(t_{yr})$ between
$8.2-8.6$.
In contrast, various authors found a significant amount of gas and dust in the 
centre of \object{NGC\,185} (Young \& Lo \cite{yolo}; Lee et al. \cite{lee};
Martinez-Delgado \& Aparicio \cite{MDA}; Martinez-Delgado et al. \cite{MDA2}).
Butler \& Martinez-Delgado (\cite{butler}) obtained an age of about $400$\,Myr for the 
youngest, centrally concentrated stars. Kang et al. (\cite{kang}) speculate that the
M-giant population in \object{NGC\,185} contains stars with a wide range of ages,
possibly representing two different epochs of star formation at
$log(t_{yr})\approx 9.0-9.4$ and $7.8-8.5$\,dex. 
In the outer parts of \object{NGC\,185} stars with ages of at least $1$\,Gyr are found. 

The red giant content of these galaxies was analysed by Nowotny et al. 
(\cite{wn}, hereafter Paper\,I). The detected AGB stars were characterised
according to the chemical properties of their atmospheres by applying an
efficient method to single out \mbox{C-rich} stars, namely, the use of
narrow-band wing-type filters that are centred around spectral molecular
features of \element[][]{TiO} and \element[][]{CN} (at
$\lambda$\,$\approx$0.8\,$\mu$m).
Within a field of view (FOV) of $6\farcm 5 \times 6\farcm 5$, the authors identified 
154 \mbox{C-rich} stars in \object{NGC\,185} and 146 \mbox{C-rich} stars in \object{NGC\,147} plus
several hundred M-Type stars on the upper giant branch in both galaxies. 
This large number of identified AGB stars motivated a search for long-period
variables in these stellar systems. An interesting aspect was if the different
metallicities and SFHs would be reflected in the PLRs of the LPVs in the two 
galaxies.

\begin{figure*}
\resizebox{\hsize}{!}{\includegraphics{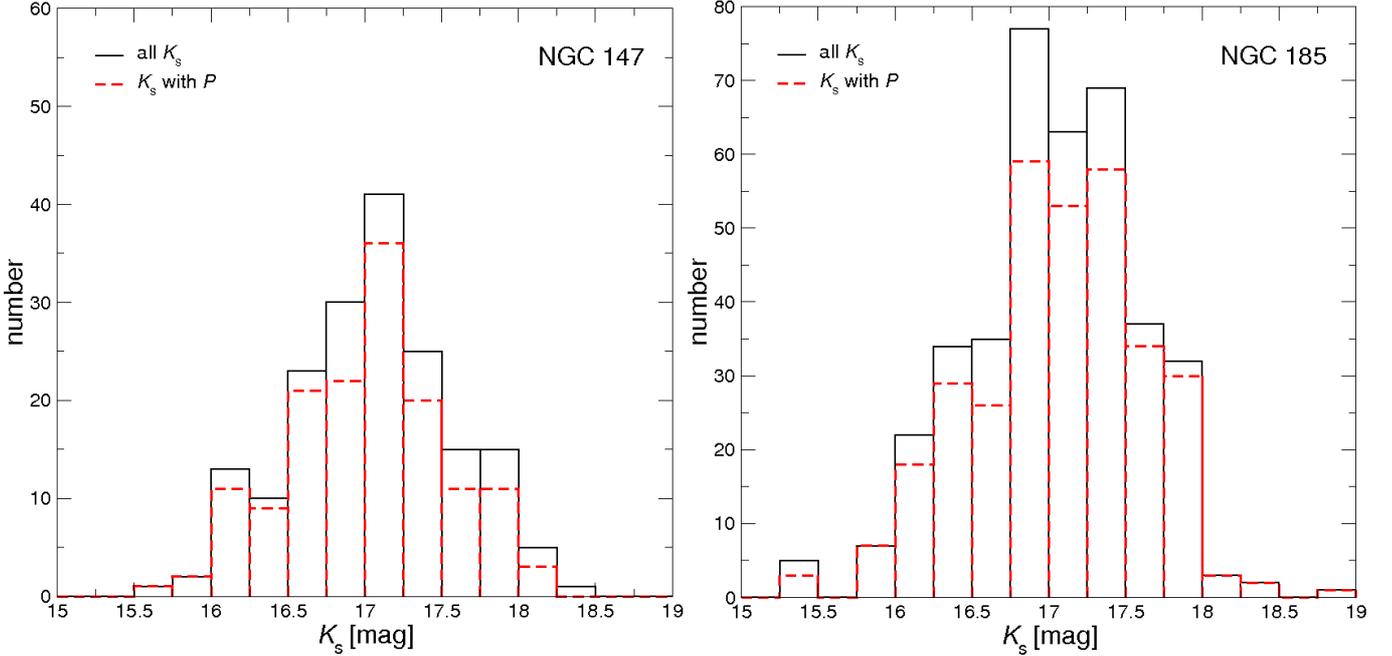}}
\caption{
$K_s$-luminosity functions of objects in \object{NGC\,147} and \object{NGC\,185}. The black line shows the 
distribution for all detected LPVs with \mbox{$K_s$-magnitudes} available (182 in
\object{NGC\,147} and 387 in \object{NGC\,185}). The red dashed line represents LPVs for which
it was additionally possible to determine a period (147 in \object{NGC\,147} and 323 in 
\object{NGC\,185}) as described in Sect.\,\ref{variables}. 
}
\label{kDist}
\end{figure*}

The main aim of this work was to identify LPVs in \object{NGC\,147} and \object{NGC\,185}.
In Sect.\,\ref{variables} we present a catalogue of red giant variables, 
the outcome of a photometric monitoring in the $i$-band. 
The results of this study are summarised in Sect. \ref{results}.

%
\section{Observations}
\subsection{Photometric monitoring}
We obtained multi-epoch observations in the $Gunn$-$i$-band with the $2.56$\,m 
Nordic Optical Telescope (NOT). The target galaxies were observed on 38 nights
in service mode between October 2003 and February 2006 with the Andalucia Faint
Object Spectrograph Camera (ALFOSC). 
It has a pixel scale of 0.19\,arcsec/pixel resulting in a FOV of approximately 
$6.4 \times 6.4$\,arcmin.
At every epoch we obtained a single image pointing towards the centre of each
galaxy. Our field covered a region corresponding to approximately one scale 
length derived from the stellar density distribution of \object{NGC\,147} 
(Battinelli \& Demers \cite{batti_b}) and \object{NGC\,185} (Battinelli \& Demers 
\cite{batti_a}), respectively.
We obtained 35 images of \object{NGC\,147} and 34 frames of \object{NGC\,185} with a sampling
period of $\approx$14 days. One example image of the time series for each of
our science targets is shown in Fig. \ref{GalLPV}. 
The time series exhibits two larger gaps of
approximately six months during which the targets were not observable.
Calibration frames to correct for sky and bias were recorded for each night of
observation. In the rare cases of missing sky flats, these were replaced by
average flats from the previous and the following observation.

\subsection{$K_s$-band photometry} 

As can be seen from spectral energy distributions of cool AGB stars (Nowotny et al. 
\cite{nowo2}), they emit most of their flux in near-infrared wavelengths. 
Hence, the $K_{s}$-band is a good measure of the bolometric flux.
The most evolved, dust-enshrouded AGB stars can be detected only at 
infrared wavelengths.
Therefore, the $K_{s}$-band has been widely used (e.g. Wood \cite{woodc}) 
to construct PLRs of LPVs. 
To allow a comparison of our results with previous studies, we
carried out single-epoch $K_{s}$-band photometry for our target systems using
NOTCam during two consecutive nights in September 2004.
This camera is equipped with a $1024 \times 1024$ HgCdTe Rockwell Hawaii 
array with a plate scale of 0.234\,arcsec/pixel resulting in a FOV of 
$4 \times 4$\,arcmin using the wide-field imaging mode of NOTCam. 
To resemble the FOV of ALFOSC, we obtained a mosaic of four partly overlapping 
dithered images per galaxy.
Accordingly, the combined FOV of the four quadrants is $\approx 6 \times 6$\,arcmin.

%
\section{Data reduction}\label{dataReduction}
\subsection{Monitoring data}
All frames obtained for this study were bias-, sky- and flatfield-corrected using 
standard data reduction routines. 
As in Paper\,I, the whole sample of stars was corrected for interstellar
reddening adopting the values from the NASA Extragalactic Database (\object{NGC\,147}: 
$A_{V}$=$0\fm574$, $A_{i}$=$0\fm336$, $A_{K_{s}}$=$0\fm064$; \object{NGC\,185}: 
$A_{V}$=$0\fm604$, $A_{i}$=$0\fm354$, $A_{K_{s}}$=$0\fm067$). 
Images taken in the $i$-band with ALFOSC also suffer from fringing. To 
compensate for this effect, it would have been necessary to obtain flatfield
images before and after each integration of the science target to
create a fringe map. Without these additional calibration images, a
correction for this effect was not possible. The maximum amplitude of variations
caused by fringing is, however, well below the minimum amplitude expected for
LPVs. 
The detection of variable stars was carried out using the image subtraction tool 
ISIS\,2.1\footnote{http://www2.iap.fr/users/alard/package.html} of Alard
(\cite{alard}). One carefully chosen $i$-band image was taken as reference frame
to obtain differences in flux relative to each image of the time
series. 
To produce light curves from these differences, we measured fluxes for each star 
on the reference frame by using a PSF fitting software written by Ch. Alard. 
Short descriptions of the code, which was originally developed for the DENIS
project, can be found in Schuller et al. (\cite{schuller}) or Beaulieu et 
al. (\cite{beau}). 
As can be seen in Fig.\,\ref{GalLPV}, the central region of \object{NGC\,185} 
is more compact towards the centre. Hence, the identification of variable stars
towards central regions is incomplete because of crowding.
The photometric zero-point correction was determined using a
sample of constant stars on the reference frame that were cross-correlated with
their counterparts in Paper\,I. To estimate a photometric error, two
samples of randomly chosen constant stars common to all images of the time
series were selected. Following the same approach as for the reference frame, 
mean zero-points were calculated from one sample of constant stars for each 
frame and subsequently used to remove zero-point variations between the various
frames. Then, the differences between the corrected magnitudes of all stars of
the second sample and the corresponding values from Paper\,I were determined. 
Their standard deviations served as an estimate for the photometric errors. The
resulting errors in the $i$-band at a mean luminosity of $19\fm5$ for the
various epochs range between $0\fm085$ for \object{NGC\,147} and $0\fm094$ for \object{NGC\,185},
respectively.

\subsection{Near-infrared data}
$K_{s}$-band images taken with the NOTCam suffer from distortion that severely 
increases towards the edge of the frame.
Thus, the frames had to be corrected for this effect before carrying out the 
standard image reduction steps.
G\r{a}lfalk (\cite{galfalk}) constructed a model of the NOTCam\,WF camera
distortion based on his observations of B335.
This NOTCam-model\footnote{http://www.astro.su.se/$\sim$magnusg/NOTCam\_ dist/} 
was implemented in a software provided by G\r{a}lfalk (\cite{galfalk}, written
in IDL, which performs additional corrections), which was used for distortion
correction of all $K_{s}$-band images.
Subsequently, the usual reduction steps of near-IR imaging were applied to the 
dithered $K_{s}$-band images. All frames belonging to one quadrant were aligned
and merged to one image to achieve a higher S/N. Point-spread functions fitting
photometry was carried out using the DAOPHOT/ALLSTAR photometry package (Stetson
\& Harris \cite{stetson}). 
The photometric zero-points to calibrate $K_{s}$ were derived using constant 
stars in each quadrant of the target galaxy, which were also found in the 2MASS
Point Source Catalogue (Cutri et al. \cite{cutri}). 
$K_s$-magnitudes of the detected variables in both galaxies are listed 
in the fifth column of Table\,\ref{periods147} and \ref{periods185},
 respectively, which are available online only.
The corresponding mean photometric uncertainties after the calibration 
are listed in Table\,\ref{errors} for different bins of magnitude within the 
range $16 < K_s < 19 $. 
In Fig.\,\ref{kDist} the
luminosity function (LF) in $K_s$ of stars detected in \object{NGC\,147} and \object{NGC\,185} 
is shown as a black continuous line. The red dashed line represents the
distribution of LPVs for which we were able to assign a period as well. 
With the instrument setting mentioned above, the photometry in $K_s$ for both
galaxies is probably complete down to $\approx$17\,mag. At a mean $K$-band luminosity
of 17 \,mag, the typical photometric errors for \object{NGC\,147} and \object{NGC\,185} are 
$0\fm15$ and $0\fm16$, respectively.

\subsection{Catalogue of variables}\label{photVar}\label{variables}
\begin{figure}
\resizebox{\hsize}{!}{\includegraphics{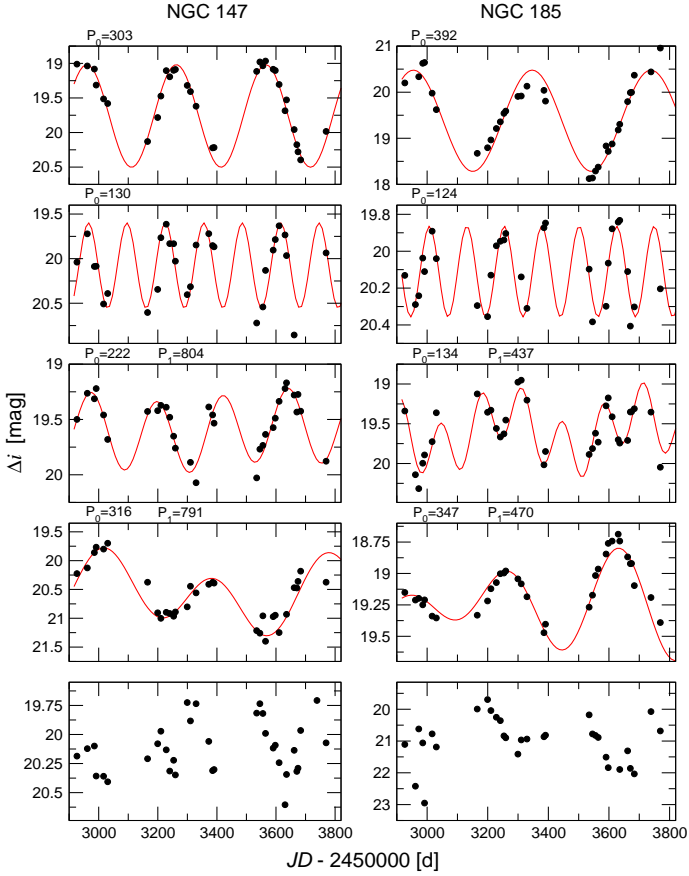}}
\caption{
Example light curves of detected LPVs in each of the target galaxies (\object{NGC\,147} 
on the left and \object{NGC\,185} on the right side). 
The black dots illustrate the observational data, while the red lines display 
the fitted light curve derived with SIGSPEC (Reegen \cite{piet}). 
Shown are different types of LPVs (mono-periodic variations, two periods, and 
LPVs with no significant period).
}
\label{LC}
\end{figure}

\begin{table}
\caption{Photometric uncertainties for the two galaxies obtained in the $i$- and $K_s$-filters.} 
\label{errors} 
\centering 
\begin{tabular}{c c c c} 
\hline\hline 
$i$ & $e_{phot}$ & $K_s$ &  $e_{phot}$ \\
\hline
$>19$ &	0.03 & $>16$ & 0.09 \\
19-20 & 0.04 & 16-17 & 0.11 \\
20-21 & 0.06 & 17-18 & 0.13 \\
21-22 & 0.09 & 18-19 & 0.18 \\ 
$<22$ & 0.16 & $<19$ & 0.26 \\
\hline 
\end{tabular}
\end{table}

The $i$-band light curves were searched for periodicities using 
SIGSPEC\footnote{http://www.sigspec.org/} (Reegen \cite{piet}). A maximum of two
periods was derived from the Fourier analysis if the criterion for significance
was fulfilled. The significance ($sig$) of a period is defined in SIGSPEC as the inverse
of the logarithmic scaled false-alarm-probability (FAP) that a discrete Fourier
transform amplitude is caused by noise (see Reegen \cite{piet} for details). 
A spectral significance of 5 therefore
corresponds to an inverse FAP of $10^5$ or, in other words, the risk of the
amplitude being just caused by noise is 1:10$^5$. Example light curves showing
different types of LPVs from both galaxies together with our best model fit 
are shown in Fig.\,\ref{LC}. The results of the period search are summarised in
Table\,\ref{periods147} and \ref{periods185}, which are available online only. 
Beside the periods and corresponding significance of the detected LPVs, 
the table also lists the mean $i$-magnitudes that were obtained from the light curve. 
The corresponding photometric errors of the mean brightness 
in the $i$-band after the calibration are listed in Table\,\ref{errors} for 
different ranges of magnitude.
We only used Fourier-amplitudes from the SIGSPEC-output to fit 
the light curves (see red line in Fig.\,\ref{LC}). Additionally, we defined a
$\sigma$-amplitude that is twice the statistical standard deviation
from the mean brightness of the variable. 
A purely sinusoidal light curve, for example, with a peak-to-peak-amplitude 
$A=1\fm0$ would result in a corresponding $\sigma$-amplitude of
$A_{\sigma}=0\fm701$ (hereafter $\Delta i$). This allows us to have a better
understanding of the overall variability of the detected LPVs in both galaxies 
even for LPVs for which no significant period could be asserted. 
In addition, this parameter is not sensitive to outliers of the observed light
curve mainly caused by dead pixels on the frame or cosmic rays during the
integration.
Depending on the results of the period search, we were able to assign one, two, or
no period to each LPV. For some stars (starting from ID\,147V000169 in 
Table\,\ref{periods147} and ID\,185V000420 in \ref{periods185}) it was not possible to 
detect a significant period, although they clearly are variable. Therefore, we
listed their $\sigma$-amplitudes $\Delta i$ to obtain
a better impression of their variability.

\subsection{Cross-correlation with photometry from Paper\,I ($V,i$,\textit{{TiO},{CN}})}
In Paper\,I, single epoch $Vi$ photometry was discussed as part of a photometric 
survey of Local Group galaxies. Furthermore, narrow-band filters (wing-type)
were used to derive information on the probable spectral types of the bright red
giant stars in these galaxies. To discuss LPVs of \object{NGC\,147} and \object{NGC\,185}
in more detail, a cross-correlation with the results obtained in Paper\,I was
performed using the DENIS software `Cross Color' written by Ch. Alard (see
Schuller et al. (\cite{schuller}). This allows us to distinguish \mbox{C-rich} LPVs
from other detected variables in our sample and to study their distribution in
consecutive diagrams. 
For approximately 75\,\% of the identified variables we could assign 
counterparts in Paper\,I. The reason for the incompleteness was threefold.
First, some stars were obviously at light minimum and, thus, too weak at the
epoch of the observations of Paper\,I. Second, we had to exclude all variables
where the cross-correlation was ambiguous because of crowding. Third, a few stars
visible on the frames studied in Paper\,I had photometric errors that were too large to
be included in the final list there.

\begin{table}
\caption{Summary of detected variables in \object{NGC\,147}, grouped according to the 
information available for different sub-samples.} 
\label{table:3} 
\centering 
\begin{tabular}{c c c c c c c c} 
\hline\hline 
LPVs & $\times$ & $\times$ & $\times$ & $\times$ & $\times$ & $\times$ & $\times$ \\
Period & & $\times$ & & & $\times$ & $\times$ & $\times$ \\
$V,i$ & & & $\times$ & & & $\times$ & $\times$ \\
\element[][]{TiO}, \element[][]{CN} & & & & & & $\times$ & $\times$ \\
$K_S$ & & & & $\times$ & $\times$ & & $\times$ \\
\hline 
objects &213&168&163&182&147&122&113\\
\hline 
\end{tabular}
\end{table}
\begin{table}
\caption{Same as Table\,\ref{table:3} for \object{NGC\,185}.} 
\label{table:4} 
\centering 
\begin{tabular}{c c c c c c c c} 
\hline\hline 
LPVs & $\times$ & $\times$ & $\times$ & $\times$ & $\times$ & $\times$ & $\times$ \\
Period & & $\times$ & & & $\times$ & $\times$ & $\times$ \\
$V,i$ & & & $\times$ & & & $\times$ & $\times$ \\
\element[][]{TiO}, \element[][]{CN} & & & & & & $\times$ & $\times$ \\
$K_S$ & & & & $\times$ & $\times$ & & $\times$ \\
\hline 
objects &513&419&381&387&323&298&229\\
\hline 
\end{tabular}
\end{table}
\begin{figure}
\resizebox{\hsize}{!}{\includegraphics{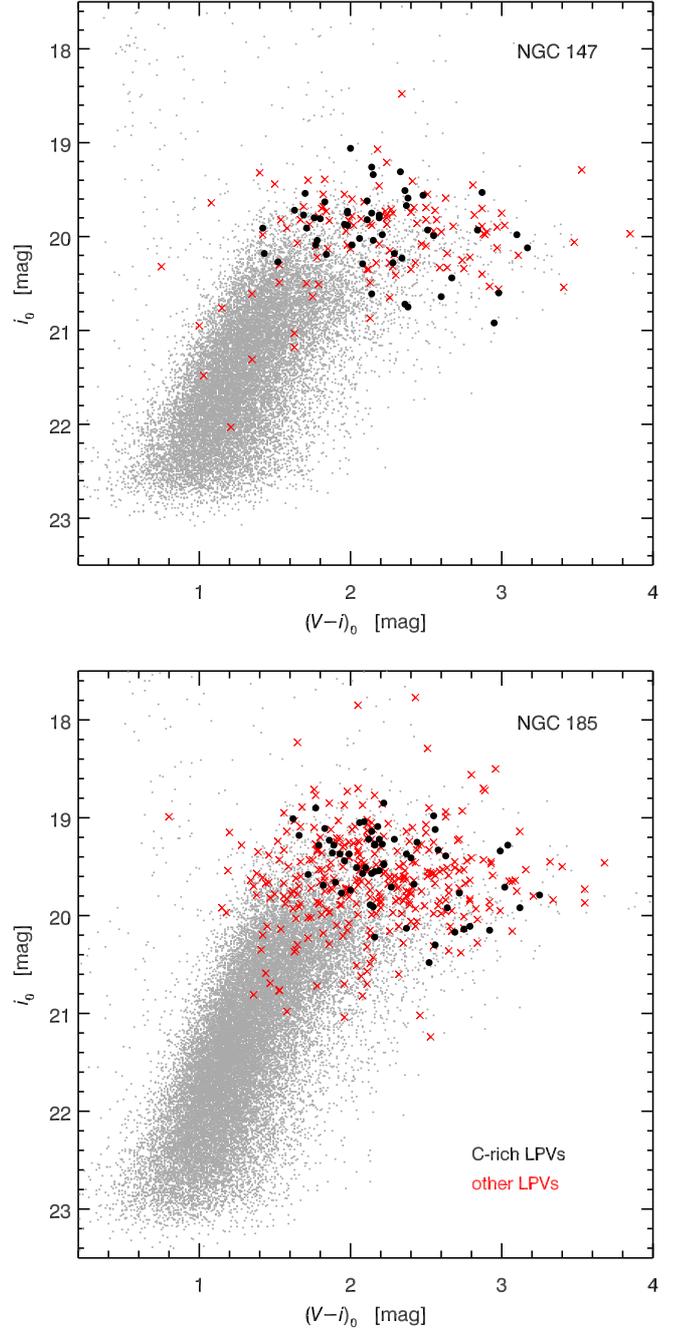}}
\caption{
Color magnitude diagrams of stars in \object{NGC\,147} (upper panel) and \object{NGC\,185} (lower 
panel) using data from Paper\,I (grey dots). Overplotted are identified LPVs of
the present study (see legend).}
\label{CMD}
\end{figure} 
%
\section{Results and discussion}\label{results}
\begin{figure}
\resizebox{\hsize}{!}{\includegraphics{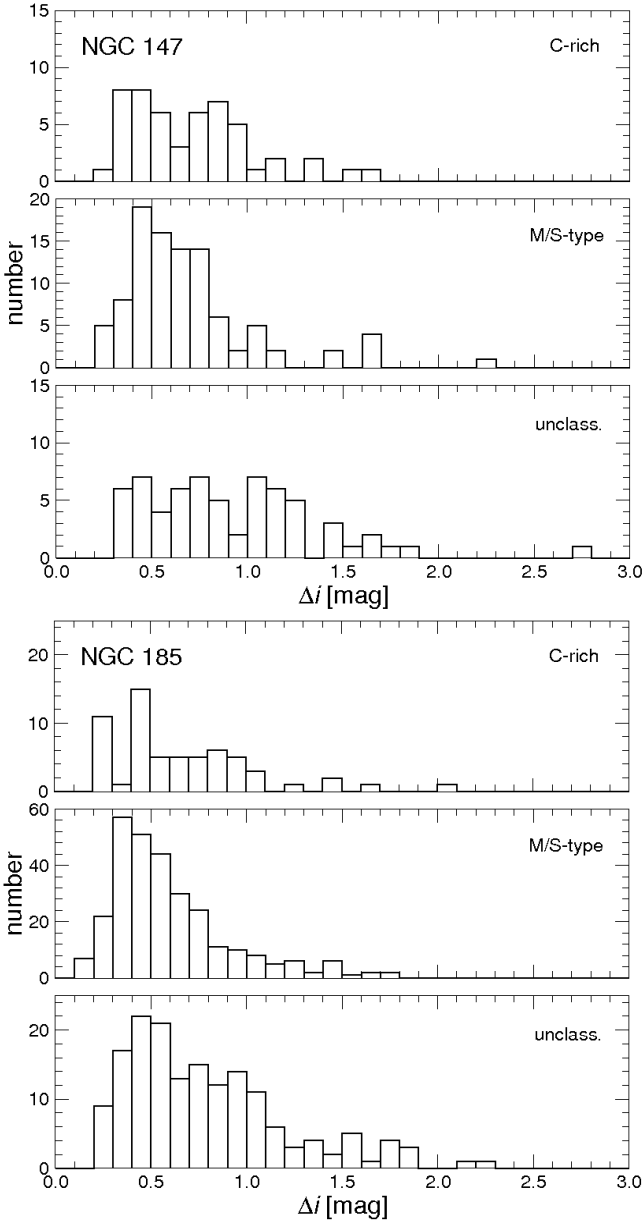}}
\caption{
Histogram of $\sigma$-amplitudes $\Delta i$ of LPVs with different surface 
chemistry in \object{NGC\,147} and \object{NGC\,185}. The first panel shows the distribution of
\mbox{C-rich}, the second panel displays \mbox{O-rich} LPVs, and the last panel
shows LPVs for which no chemical information was available (mainly because of missing narrow-band photometry).
}
\label{aHist}
\end{figure} 

\subsection{Variable stars}\label{LPVs}
In Tables\,\ref{table:3} and \ref{table:4} we list the number of objects in each 
of the two galaxies for which certain sets of data are available. We thereby
describe the size of several sub-samples grouped according to the information
available. For the 163 variables in \object{NGC\,147} and the 381 objects in \object{NGC\,185}
where we have broad-band photometry from Nowotny et al. (\cite{wn}), the
locations in the CMDs are shown in Fig.\,\ref{CMD} together with the full
sample of Paper\,I. The LPVs are superimposed as red crosses and those
classified as \mbox{C-rich} stars (according to Paper\,I) are drawn as black
circles. Evidently, most of the identified LPVs are located in the
upper region of the giant branch where AGB stars, and in particular classical
\mbox{carbon-rich} stars, are expected. Approximately two thirds of the variable 
red giant stars in both galaxies clearly show mono-periodic light variations. 
Two periods could be assigned to 20
variables of \object{NGC\,147}, and for 45 LPVs of this system no significant period was
found. In \object{NGC\,185} we find 38 LPVs exhibiting two periods and 94 LPVs for which
no period could be detected. The light curves of LPVs without a significant
period show a wide variety of shapes. A part of them definitely shows very long
variations that exceed the length of our time series. On the other hand, we also
have a short period limit because of the sampling interval of our observations,
which amounts to approximately 90 days. We also found objects with irregular
light variations, sometimes alternating with phases of comparably constant
brightnesses.
A selection of different LPV light curves detected in both galaxies is presented
in \mbox{Fig.\,\ref{LC}}. Mono-periodic cases can be found in the panels of
the upper two rows.
Examples of LPVs exhibiting two significant periods are plotted in the following 
two rows, which show a beating phenomenon in the second panel of the fourth row.
In the last row of \mbox{Fig.\,\ref{LC}}, two cases out of the $\approx 20\%$ 
of our sample stars are given for which no significant period could be 
determined from the observations. 
However, taking into account their $(V-i)$, $\sigma$-amplitudes 
and the time scales of their light variations, they can still clearly be 
classified as LPVs.
\begin{table}
\caption{LPVs identified in the two target galaxies grouped according to their 
chemical types as derived from the narrow-band photometry of Paper\,I.} 
\label{table:LPVsubsamples} 
\centering 
\begin{tabular}{c c c} 
\hline\hline 
& \object{NGC\,147} & \object{NGC\,185} \\
\hline
C & 51 & 61 \\
M/S & 98 & 288 \\
unclass. & 64 & 164 \\
\hline 
$\Sigma$ & 213 & 513 \\
\hline 
\end{tabular}
\end{table}

Using the narrow-band photometry from Paper\,I, we can assign a probable 
atmospheric chemistry to most of the LPVs in our sample, and study the
variability characteristics in relation to these defined subgroups. Of special
interest are carbon stars, which we assume to be intrinsic post-third-dredge-up
objects. Table\,\ref{table:LPVsubsamples} groups the identified variable stars
according to the designated chemistry type. To search for possible 
correlations between the location of LPVs within the galaxy and their chemistry, 
we chose one of the CCD images from the \mbox{$i$-band} time series and indicated
the detected variables as red circles and \mbox{C-rich} LPVs as blue 
circles (see Fig.\,\ref{GalLPV}).
The radial distribution for identified AGB stars in \object{NGC\,147} and \object{NGC\,185} has 
been discussed in Paper\,I. The authors find similar distributions for all AGB 
stars and for only \mbox{C-rich} AGB stars. 
If the sample is reduced to detected LPVs in those galaxies, the
trend in radial distribution for \mbox{C-rich} LPVs is similar as that for the full 
sample of LPVs.
The number ratio of carbon-rich over
oxygen-rich LPVs amounts to 0.52 for \object{NGC\,147} and 0.21 for \object{NGC\,185},
respectively. These values fall between the corresponding ratios for the whole
population and the ratios when limiting the O-rich sample to spectral types M5
and later, as presented in Paper\,I. Thereby, our sample selected on variability
provides a good representation of the AGB population with which the LPV class
is typically associated. If we look at the histograms in Fig.\,\ref{aHist}, we
notice that the amplitude distributions of the O-rich objects is dominated by
small-amplitude variables. Indeed, the histograms suggest that the peak at
$0\fm35$ is not real, but that we are missing stars with the shortest amplitudes
owing to the limited sampling rate of our monitoring. In contrast, C-rich
stars exhibit a much flatter distribution in these plots. For the sake of
completeness, the distribution of LPVs without narrow-band photometry is given
in the last panels of Fig.\,\ref{aHist} for each of the galaxies. 
\begin{figure}
\resizebox{\hsize}{!}{\includegraphics[angle=-90]{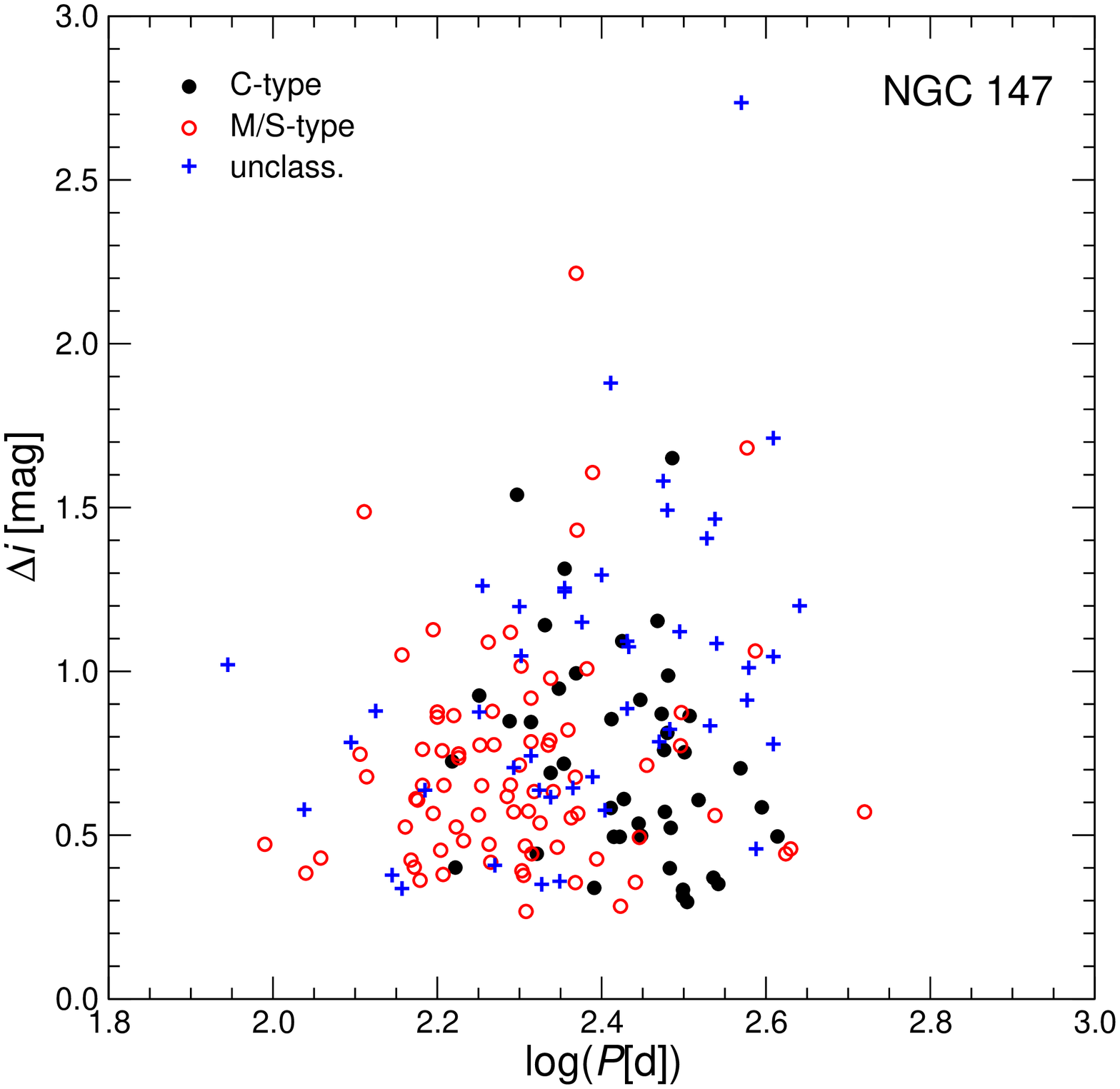}}
\resizebox{\hsize}{!}{\includegraphics[angle=-90]{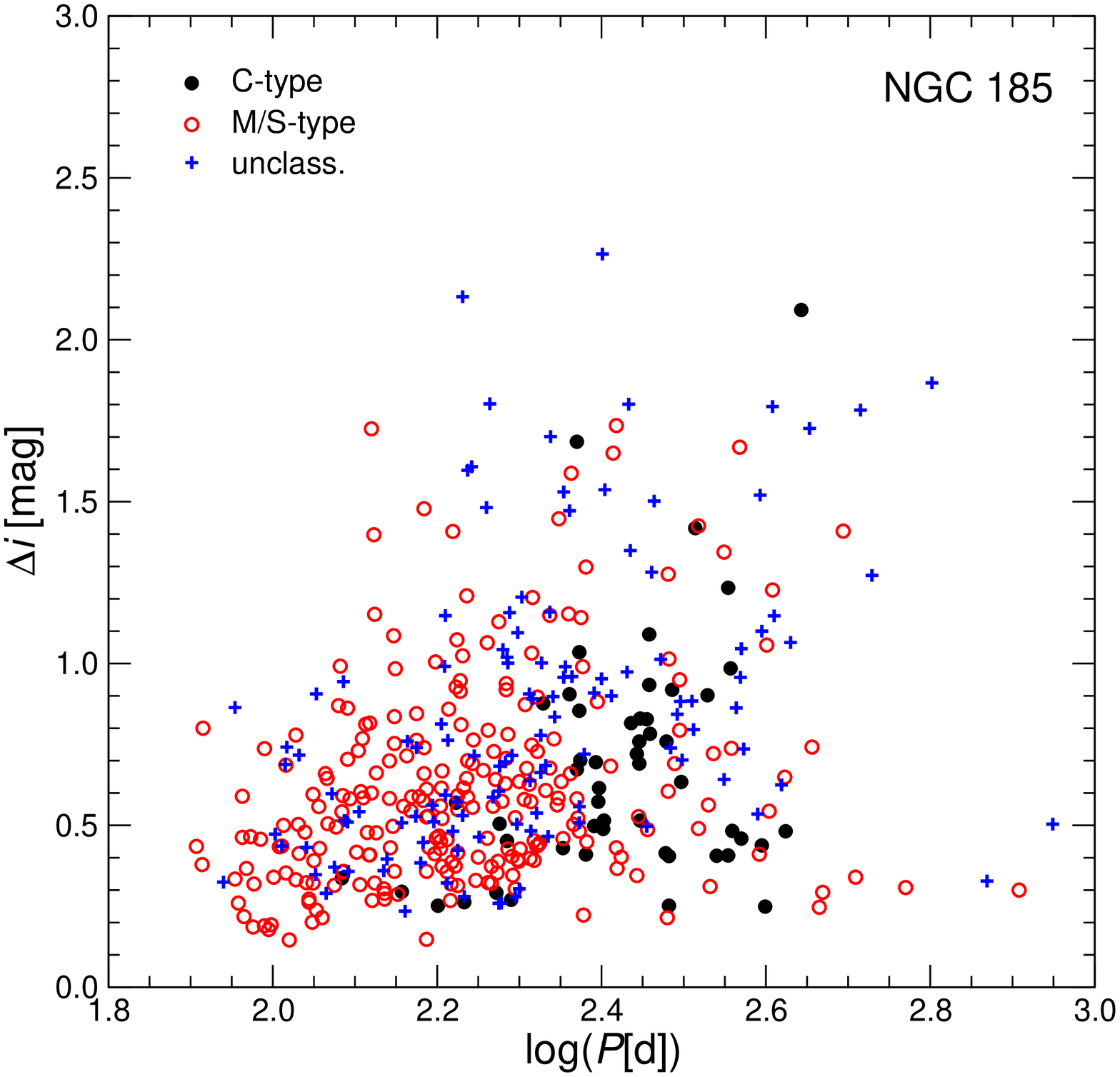}}
\caption{
Photometric amplitudes $\Delta i$ versus periods for LPVs identified in
\object{NGC\,147} (upper panel) and \object{NGC\,185} (lower panel).
\mbox{C-rich} objects are indicated by black filled circles,
while \mbox{M/S-type} stars are plotted with red open circles,
and unclassified LPVs are drawn as blue crosses.
}
\label{lgPAmpl2}
\end{figure} 
The period distributions of the LPVs in the two galaxies can be seen in the 
$\Delta i$ vs. log\,$P$ diagrams in Fig.\,\ref{lgPAmpl2}. Here, we plotted only
the first significant period of all detected LPVs. The range of periods covered
by the variables is similar in both systems ($\approx$\,$90$--$600^{\rm d}$) and a
weak tendency of larger amplitudes with increasing period may be visible.
Considerably more LPVs with shorter periods were found in \object{NGC\,185}. In this
galaxy, we also found a small group of stars with very long periods and small
amplitudes. These are likely candidates for LSPs, but no other significant
periodicities were found from our times series. Splitting up the stars according
to their chemistry reveals a concentration of C-rich targets around $\log
P$\,=\,2.5, while the O-rich stars are found predominantly at shorter periods.
This behaviour is expected from theory.

\begin{figure}
\resizebox{\hsize}{!}{\includegraphics[clip]{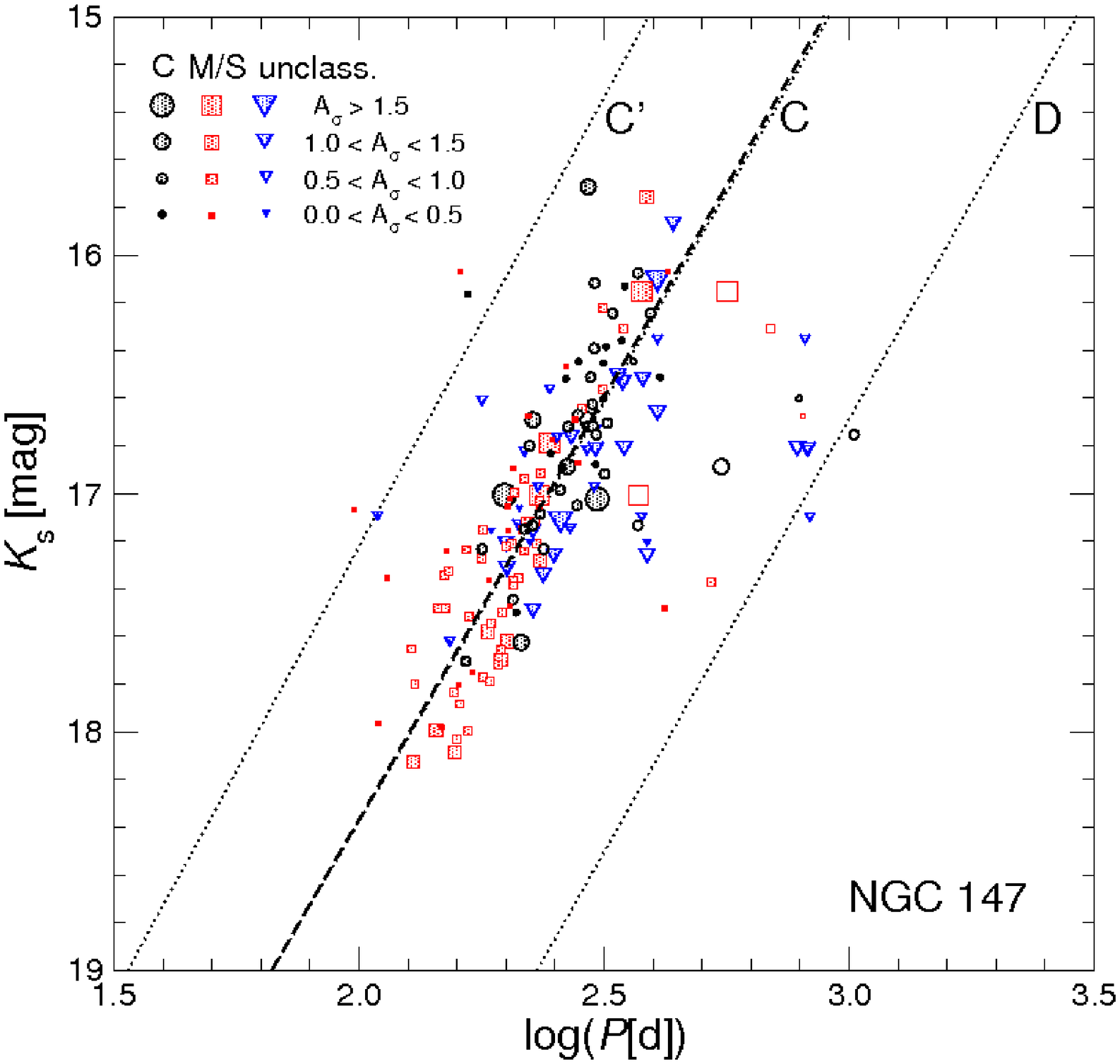}}
\resizebox{\hsize}{!}{\includegraphics[clip]{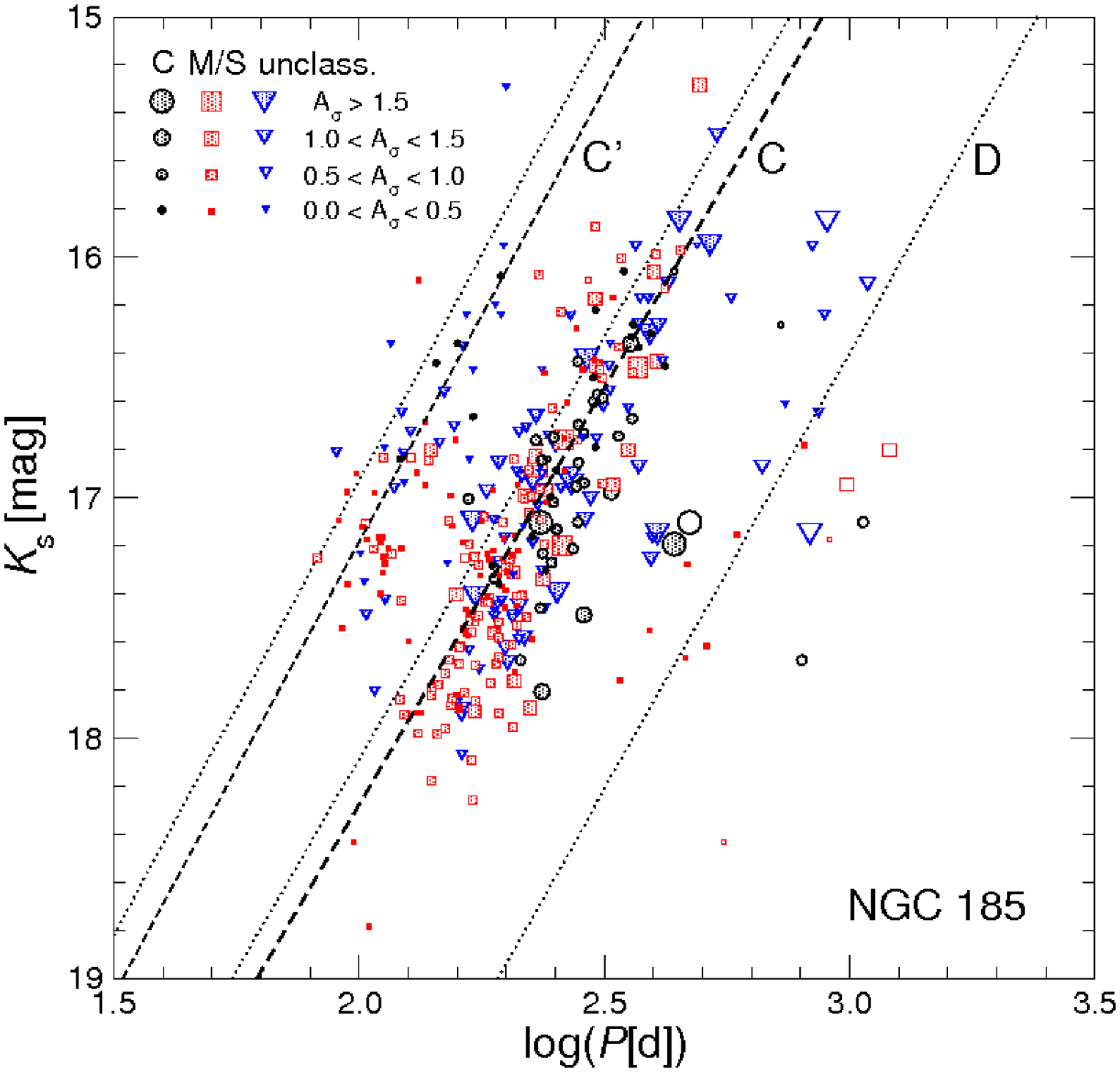}}
\caption{
Period-$K_s$-diagram of \object{NGC\,147} (upper panel) and \object{NGC\,185} (lower panel). 
\mbox{C-type} LPVs are plotted as black circles, \mbox{M/S-type} LPVs as red
squares and unclassified variable red giant stars as blue triangles. 
The sizes of the plot
symbols are scaled corresponding to the photometric amplitudes $A_{\sigma}$ (see
legend). For LPVs with two detected periods, open symbols indicate the second,
longer period of both. The black dotted lines (C$^{\prime}$, C, D) mark the PLRs
found for the LMC by Ita et al. (\cite{ita}). They were adopted and shifted
according to the distance moduli differences between the LMC and our galaxies
(see text). The black thick dashed line shows the PLRs derived by using the LPVs
identified in this work. The black thin dashed line in the lower panel is plotted 
for demonstration purpose only (see text).}
\label{KlgP}
\end{figure}
\subsection{Period-luminosity relations}\label{PLR}
To construct period-luminosity diagrams (PLDs) for both target systems, \object{NGC\,147} 
and \object{NGC\,185}, the datasets of the $i$-time series and the $K_{s}$-band
photometry were combined. This resulted in 182 LPVs with $K_{s}$-magnitudes and
detected periods in \object{NGC\,147} and 387 LPVs in \object{NGC\,185}, respectively, which could
be used for the construction of the PLDs. The resulting $K_{s}$-$\log{P}$
diagrams for both galaxies are shown in Fig.\,\ref{KlgP}. Different symbols
denote the various classes of LPVs, namely \mbox{C-rich}, \mbox{O-rich}, and 
unclassified variables according to the narrow-band photometry adopted from 
Paper\,I. Furthermore, the amplitude was used to group the variables into four
sub-samples indicated by the symbol sizes in Fig.\,\ref{KlgP}. 
Obviously, most LPVs in both galaxies seem to form a distinct sequence at 
the very same location in the PLDs as the sequence of fundamental mode pulsators
(labeled sequence~C; cf. Sect.\,1) found by various authors for the Magellanic Clouds.
For illustration purposes, we overplotted the relations of Ita et al.
(\cite{ita}). To shift their relations according to the difference in distance
between the Magellanic Clouds and our galaxies, we adopted the distance moduli
determined via the brightness of horizontal branch stars for our target systems
by Butler et al. (\cite{butler}). They derived distance moduli of $(m -
M)_{0}$=$24\fm38\pm 0\fm01$ and $24\fm09\pm0.06$~mag for \object{NGC\,147} and \object{NGC\,185},
respectively. For the LMC we adopted the distance modulus obtained by
Pietrzy\'{n}sky et al. (\cite{pietr}) of $(m - M)_{0}$=$18\fm50\pm 0\fm06$.
With respect to the atmospheric chemistry, we find O-rich stars along the whole 
sequence (with a slight thinning in number towards the top), while C-rich stars
mainly occupy the upper part of the sequence. This agrees with findings
of other studies (e.g., Wood \cite{woodc} or Ita et al. \cite{ita}). 
After applying a 3.0\,$\sigma$ clipping to exclude stars that are considered not 
to belong to sequence~C, a least-squares fitting was performed to obtain PLRs
for this sample of stars. 
The linear regression $(m_{K}$\,=\,$a \log{P} +
b)$; black dashed lines in Fig.\,\ref{KlgP}) of this selection resulted in a
slope $a$ of $-3.55$ for \object{NGC\,147} and $-3.47$ for \object{NGC\,185}. For the 
\mbox{intercepts\,$b$} of each relation we obtained 25$\fm$46 and 25$\fm$22 
for \object{NGC\,147} and \object{NGC\,185}, respectively. 
In Table\,\ref{PLRs} we contrast PLRs derived by different studies
with those found here. This comparison of results for different stellar
systems is of interest for studying the aspect of the universality of the PLRs
of Sequence~C stars, which may then serve as an additional tool to measure distances in
extragalactic systems. 
Owing to the limited number of carbon stars in our samples,
we did not analyse \mbox{C-rich} and \mbox{O-rich} stars separately. 
However, for comparison reasons we calculated PLRs for 
our samples with and without \mbox{C-rich} LPVs. 
The small differences of the values demonstrate that the results are robust against 
slight changes of the sample selection.
For additional calculations we only consider the results of the complete 3.0\,$\sigma$ 
clipping sample.
The slopes $a$ found for
\object{NGC\,147} and \object{NGC\,185} are close to those given for the combined samples in
other stellar systems. 

In Fig.\,\ref{KlgP} an obvious offset between the shifted 
LMC-PLRs and
those derived for \object{NGC\,185} can be seen. The numbers in Table\,\ref{PLRs} suggest
that this shift amounts to approximately $0\fm 2-0\fm4$ (depending on sample selection
and regression method). A mild difference in
the zero-point $b$ is expected owing to the difference in metallicity (Wood
\cite{wooda}). However, for the lower metallicity of \object{NGC\,185} relative to the
LMC, a star should be brighter at a given period, making the discrepancy even
larger. A simple but not necessarily final explanation for this difference
would be an error in the distance modulus of \object{NGC\,185}. 
When attempting to bring our observations in line with the LMC relations 
of Ita et al. (\cite{ita}), a distance modulus of 24$\fm$30 seems to be more
appropriate for this galaxy. This value is obtained by subtracting the 
resulting $K_{s}$-magnitudes derived from the relation of Ita et al. (\cite{ita}) and 
from this work at a constant value of $\lg P = 2.31 $. 
For \object{NGC\,147}, the distance modulus from the literature excellently agrees 
with our data. The zero-point problem for \object{NGC\,185} needs further exploration. 
However, if the reason for the offset PLRs is indeed an error in the distance
modulus, this would be the first correction of a distance to a galaxy based on
PLRs of LPVs. 

For \object{NGC\,185} there are indications of another parallel sequence of
LPVs shifted towards shorter periods (sequence~C$^{\prime}$). On average, stars
on this sequence exhibit smaller amplitudes than objects on sequence~C (see
lower panel of Fig.\,\ref{KlgP}). A similar PLR was found in the LMC (e.g. Ita
et al. \cite{ita}, Fraser et al. \cite{fraser2}) and is associated with first
overtone pulsation. The smaller light amplitude of this group identified in our
sample agrees with this interpretation. 
Note that the trend for the zero point of sequence~C is also visible for
sequence~C$^{\prime}$ but the number of detected LPVs populating sequence~C$^{\prime}$ 
in \object{NGC\,185} prohibits another linear regression. 
For demonstration purpose only, we plotted sequence~C$^{\prime}$ of Ita et 
al. (\cite{ita}) and shifted it according to the difference in distance obtained
for sequence~C. This line is drawn as thin dashed line in Fig.\,\ref{KlgP}.
A handful of targets in \object{NGC\,147} may also be located on this sequence.
Variables corresponding to higher overtone pulsation were not accessible to our 
study owing to the sampling rate limitations. 

From studies in various stellar
systems it is known that the LSPs, visible in a significant fraction of all
LPVs, form another sequence to the right of sequence~C in the PLD. This
sequence~D, taken from Ita et al. (\cite{ita}), is overplotted in both panels of
Fig.\,\ref{KlgP} as well. Evidently, we found objects in our variability
study for which long secondary periods derived from the light curve seem to
cluster around this sequence~D. In these cases, the primary period is typically
located close to the fundamental mode pulsation sequence. Note that a detailed 
determination of these LSPs in our sample is hampered by the limited length of 
our time series. As described in Sect.\,\ref{s:intro}, the interpretation of
this kind of variability is still a matter of debate. Picking up the results of
Wood \& Nicholls (\cite{wooni}) that LSP-stars show a significant mid-infrared
excess (circumstellar dust), the corresponding targets from our monitoring could
be expected to be promising candidates for detecting signatures of circumstellar
material.

\begin{table}
\begin{center}
\caption{Comparison of our $P$-$L$ relations with literature values for different 
stellar systems.
All relations were shifted to an absolute scale $M_{K}$ using the distance
moduli given in the text.}
\begin{tabular}{l ccccc}
\hline
\hline
          & \multicolumn{2}{c}{$M_{K_s}=a \log P + b$}&\\
          &  $a$  &  $b$ & ($m$--$M$) & Refs.\\
\hline
Galactic  &--3.56 $\pm$ 0.17     & 1.14 $\pm$ 0.42     &            &1\\
 LMC all  &--3.52 $\pm$ 0.03     & 1.04 $\pm$ 0.08     & 18.50      &2\\
 LMC      &                      &                     &            & \\
 all      &--3.34 $\pm$ 0.02     & 0.40 $\pm$ 0.05     & 18.50      &3\\
O-rich    &--3.31 $\pm$ 0.04     & 0.47 $\pm$ 0.09     & 			&3\\
C-rich    &--3.16 $\pm$ 0.04     & --0.10 $\pm$ 0.11   &            &3\\
LMC       &                      &                     &            & \\
all   &--3.57 $\pm$ 0.16     & 1.20 $\pm$ 0.39     & 18.50      & 4\\
O-rich     &--3.47 $\pm$ 0.19     & 0.98 $\pm$ 0.45     &            & 4\\
C-rich     &--3.30 $\pm$ 0.40     & 0.48 $\pm$ 0.98     &            & 4\\
 Cen\,A   &--3.37 $\pm$ 0.11     & 0.80 $\pm$ 0.29     & 27.87      & 5 \\
\hline
NGC\,147  &                      &                     &            & \\
all 	  &--3.55 $\pm$ 0.15 & 1.08 $\pm$ 0.36 & 24.38 &6\\
O-rich    &--3.81 $\pm$ 0.25 & 1.68 $\pm$ 0.58 & 24.38 &6\\
NGC\,185  &                      &                     &            & \\
all 	  &--3.47 $\pm$ 0.11 & 1.13 $\pm$ 0.27 & 24.09 &6\\
O-rich    &--3.72 $\pm$ 0.16 & 1.68 $\pm$ 0.38 & 24.09 &6 \\
\hline
\end{tabular}
\label{PLRs}
\end{center}
\tablebib{(1)~Groenewegen \& Whitelock (\cite{groene3}); (2) Ita et al. 
(\cite{ita}); (3)~Riebel et al. (\cite{riebel}); (4)~Feast et al. (\cite{feast}); 
(5)~Rejkuba (\cite{rejk}); (6)~this work.}
\end{table} 

\subsection{A hidden link to the star-formation history?}\label{SFH}
The similar datasets for the two galaxies with comparable properties (cf. 
Sect.\,\ref{s:intro}) encouraged us to make a detailed comparison of the derived
PLDs (Fig.\,\ref{KlgP}). Besides the different number of LPVs detected
(Table\,\ref{table:LPVsubsamples}), which is likely related to the different
masses of the systems, the most obvious distinction is the fraction of luminous
stars found along sequence~C$^{\prime}$. For \object{NGC\,185} roughly 10\,\%
of all LPVs can be attributed to this sequence for first overtone pulsation,
while only less than 3\,\% of these stars were detected in \object{NGC\,147}. This raises
the question whether some fundamental differences between the two galaxies are
mirrored in the recognised discrepancy. A possible interpretation of the lack of
stars on sequence~C$^{\prime}$ in \object{NGC\,147} could involve a difference in the
mass distribution of these objects. Linear pulsation models (Fox \& Wood
\cite{FoxW82}) as well as observational results from LPVs in stellar clusters
(Lebzelter \& Wood \cite{lzwood} and \cite{lzwood07}) suggest an evolutionary
path of an AGB star through the PLD starting on an overtone sequence and later,
at higher luminosities, switching to the fundamental mode sequence. Because there
are variables in \object{NGC\,185} on sequence~C$^{\prime}$ with the same luminosity as
the bulk of the stars along sequence~C, this points towards a higher mass of
these stars. 

How can this difference in the mass distribution be explained?
The most obvious approach would be to assume that the two galaxies differ in
their SFH. Indications for this were found in previous studies and become
apparent, for example, in the SFH diagrams of Mateo (\cite{mateo}; Fig.\,8).
Hints for a recent star-formation episode can be found for \object{NGC\,185}, namely
a small population of younger stars concentrated in the central regions, a
significant amount of interstellar gas, and prominent dust patches
(Sect.\,\ref{s:intro} and Paper\,I). On the other hand, \object{NGC\,147} seems to be
free of dust and gas, and there are no indications for a population younger than
1 Gyr (Han et al. \cite{han}, Paper\,I).
Another hint in this direction may be the possible detection of a small shift 
in the light amplitude distribution of the LPVs (Fig.\,\ref{aHist}) in the two
systems. A younger system is expected to contain more stars with higher
masses and, thus, smaller amplitudes (Lebzelter \& Wood \cite{lzwood07})
compared to older systems. 

Theory predicts a linear trend between
the mean metallicity and the ratio of C-type to M-type stars of a galaxy, which has been 
confirmed by observations (Iben \& Renzini, \cite{iben}; Mouhcine \& Lan\c{c}on \cite{mou}).
For systems with lower mean metallicities the production of C-type stars is favoured. 
According to the values
obtained in Paper\,I, \object{NGC\,185} is considered as the more metal-poor galaxy. However,
the number of C-type stars is approximately the same in both systems, which leads to a
much lower C/M ratio for \object{NGC\,185} than for \object{NGC\,147} (0.21 and 0.52, respectively). 
Note that one has to be careful 
with the interpretation of this result because Battinelli \& Demers (\cite{batti})
clearly demonstrated how severely the C/M ratio depends on the selection criterion 
(see their Figs. 3 and 4). In addition, the mean metallicity of a galaxy is an 
elusive parameter because galaxies consist of multiple populations with a mixture of
ages and metallicities. The small separation (within the uncertainties) of our target 
galaxies in Fig.\,3 of Battinelli \& Demers (\cite{batti}) does not allow us to draw 
any conclusions.

The SFHs of the two galaxies should be explored in more detail to arrive at a 
final interpretation of this question. Because both galaxies, \object{NGC\,147} and \object{NGC\,185},
are comparable in properties (distance, luminosity, C-star content) but differ
in their SFHs, they appear to be ideal candidates to shed light on this challenging
topic. 
%
\section{Conclusions}
A photometric monitoring in the $i$-band of the two Local Group dwarf galaxies 
\object{NGC\,147} and \object{NGC\,185} led to the identification of 213 and 513 long-period
variables, respectively. Narrow-band photometry adopted from Paper\,I allowed us to
investigate the number of C-rich and O-rich stars among this variability class.
Thus, our study is one of the few (e.g., Groenewegen \cite{Groen04}) that uses
a more elaborated chemistry separation than the often used broad-band colour
criterion (e.g., ($J$\,--\,$K$)\,$>$\,1.4).
Because the attribute of long-period variability is more significant than a pure 
brightness limit to select the AGB stars among all late-type giants of a
population, the ratio of C-/M-type LPVs is a more reliable measure of the
corresponding ratio on the AGB. From our study we determine a value of 0.52 for
\object{NGC\,147} and 0.21 for \object{NGC\,185}. Our substantial sample of LPVs allowed us to
investigate the corresponding period-luminosity relations in the
$K_{s}$-$\log{P}$-plane as well. Most variables in both galaxies are located along
sequence\,C, where fundamental mode pulsators are theoretically expected.
A linear regression $(m_{K}$\,=\,$a \log{P} + b)$ was fitted to the data of 
these stars. The resulting fit parameters agree well with the
corresponding values found for the LMC in the literature (e.g., Ita et al.
\cite{ita}). This allows us to speculate further about the universality of the
P-L-relation of LPVs. However, we noticed a discrepancy in $b$ for \object{NGC\,185},
which may point to an error in the previously derived distance modulus of this
galaxy. This would be the first correction of a distance to a galaxy based on
PLRs of LPVs. 
The most significant difference between the PLDs of the two target systems is 
the presence of a group of first overtone pulsators in \object{NGC\,185}, which is almost
completely missing in \object{NGC\,147}. We speculate that this effect may be explained
by a difference in the star-formation history and, accordingly, a different mass
distribution on the AGB. According to our sampling interval, periods shorter
than 90 days could not be determined. We intend to extend our monitoring by
obtaining more observations at a higher sampling rate, which will allow the detection
of these periods. Moreover, these additional data points would increase the
chance to properly describe the long secondary periods that are likely present in a
significant fraction of our LPV sample.
\begin{acknowledgements}
The data presented here have been taken using ALFOSC, which is owned by the 
Instituto de Astrofisica de Andalucia (IAA) and operated at the Nordic Optical
Telescope under agreement between IAA and the NBIfAFG of the Astronomical
Observatory of Copenhagen.
This work was supported by the \textit{Fonds zur F\"orderung der 
Wis\-sen\-schaft\-li\-chen For\-schung} (FWF) under project numbers P18939--N16,
P20046--N16 and P21988-N16.
This research has made use of the NASA/IPAC Extragalactic Database (NED) which 
is operated by the Jet Propulsion Laboratory, California Institute of
Technology, under contract with the National Aeronautics and Space
Administration. 
\end{acknowledgements}

\Online

\begin{appendix}
\section{}
\begin{longtable}{lcccccccccccc}
\caption{\label{periods147} Period search results of \object{NGC\,147}. 
This table  
lists in the first five columns the internal IDs, the coordinates (J2000), 
mean \mbox{$i$ magnitudes} (obtained from the time series) as well as $K_s$-magnitudes.
Periods, $\sigma$-amplitudes, and the SIGSPEC-significance are found in the next three columns. 
Long-period variables exhibiting two significant periods have entries in the ninth and tenth columns listing the 
second period and its SIGSPEC-significance. The last two columns give the types 
of different chemistry and the IDs carried out by the authors of Paper\,I.
}\\
\hline\hline
ID & RAJ2000 & DEJ2000 & $i$ & $K_s$ & $P_{0}$ & $\Delta i$ & $sig_{0}$ & $P_{1}$ & $sig_{1}$ & type & ID$_{PaperI}$\\
& h:m:s & d:m:s & [mag] & [mag] & [d] & [mag] & & [d] & \\
\hline
\endfirsthead
\caption{continued.}\\
\hline\hline
ID & RAJ2000 & DEJ2000 & $i$ & $K_s$ & $P_{0}$ & $\Delta i$ & $sig_{0}$ & $P_{1}$ & $sig_{1}$ & type & ID$_{PaperI}$\\
& h:m:s & d:m:s & [mag] & [mag] & [d] & [mag] & & [d] & \\
\hline
\hline
\endhead
\hline
\endfoot
147V000001 & 00 33 30.99 & 48 32 18.07 & 20.06  & ---    & 241 & 1.01 & 2.1 & ---  & --- & S &  00971672  \\  
147V000002 & 00 33 28.57 & 48 29 33.85 & 20.20  & ---    & 258 & 0.85 & 4.6 & ---  & --- & C &  02190795  \\  
147V000003 & 00 33 27.59 & 48 29 48.00 & 20.47  & ---    & 206 & 0.74 & 5.3 & ---  & --- & u &  ---       \\  
147V000004 & 00 33 27.66 & 48 30 33.00 & 19.70  & 16.66  & 300 & 0.57 & 5.7 & ---  & --- & C &  02691110  \\  
147V000005 & 00 33 27.19 & 48 29 29.93 & 19.91  & ---    & 196 & 0.71 & 5.6 & ---  & --- & u &  02920773  \\  
147V000006 & 00 33 26.68 & 48 29 16.68 & 19.88  & 17.30  & 211 & 0.54 & 4.6 & ---  & --- & M &  03180702  \\  
147V000007 & 00 33 26.43 & 48 30 38.81 & 20.60  & 16.55  & 314 & 0.87 & 4.3 & ---  & --- & S &  03351141  \\  
147V000008 & 00 33 25.15 & 48 32 50.78 & 20.12  & 17.24  & 231 & 0.55 & 4.5 & ---  & --- & M &  04081845  \\
\end{longtable}
%
\begin{longtable}{lcccccccccccc}
\caption{\label{periods185} Same as Table \ref{periods147} but for \object{NGC\,185}.
}\\
\hline\hline
ID & RAJ2000 & DEJ2000 & $i$ & $K_s$ & $P_{0}$ & $\Delta i$ & $sig_{0}$ & $P_{1}$ & $sig_{1}$ & type & ID$_{PaperI}$\\
& h:m:s & d:m:s & [mag] & [mag] & [d] & [mag] & & [d] & \\
\hline
\endfirsthead
\caption{continued.}\\
\hline\hline
ID & RAJ2000 & DEJ2000 & $i$ & $K_s$ & $P_{0}$ & $\Delta i$ & $sig_{0}$ & $P_{1}$ & $sig_{1}$ & type & ID$_{PaperI}$\\
& h:m:s & d:m:s & [mag] & [mag] & [d] & [mag] & & [d] & \\
\hline
\endhead
\hline
\endfoot
185V000001 & 00 39 14.67 & 48 20 40.97 & 19.68   & ---    & 139 & 0.43 & 4.0 &  --- & --- &  S & 01961242 \\   
185V000002 & 00 39 14.42 & 48 18 37.24 & 19.84   & ---    & 114 & 0.43 & 4.4 &  --- & --- &  S & 02020584 \\   
185V000003 & 00 39 14.12 & 48 21 31.49 & 20.03   & ---    & 153 & 0.66 & 2.8 &  --- & --- &  M & 02281511 \\   
185V000004 & 00 39 13.29 & 48 20 28.43 & 19.58   & ---    & 147 & 0.55 & 5.7 &  --- & --- &  M & 02681175 \\   
185V000005 & 00 38 52.03 & 48 20 10.21 & 19.28   & 16.31  & 371 & 0.46 & 6.0 &  --- & --- &  C & 13961070 \\   
185V000006 & 00 39 13.12 & 48 19 25.54 & 19.57   & 17.48  & 163 & 0.46 & 5.6 &  --- & --- &  S & 02740840 \\   
185V000007 & 00 39 12.54 & 48 17 54.78 & 20.41   & ---    & 311 & 0.84 & 6.1 &  --- & --- &  u & 02990357 \\   
185V000008 & 00 39 12.44 & 48 21 06.86 & 20.41   & ---    & 123 & 0.86 & 3.8 &  --- & --- &  M & 03161379 \\
\end{longtable}
\end{appendix}


\begin{thebibliography}{}
\bibitem[2000]{alard} Alard, C. 2000,
\aaps, 144, 363 
\bibitem[2004a]{batti_a} Battinelli, P., \& Demers S. 2004,
\aap, 417, 479
\bibitem[2004b]{batti_b} Battinelli, P., \& Demers S. 2004,
\aap, 418, 33
\bibitem[2005]{batti} Battinelli, P., \& Demers S. 2005,
\aap, 434, 657
\bibitem[2008]{beau} Beaulieu, J. P., Carey, S., Ribas, I., \& Tinetti, G. 2008,
\apj, 677, 1343
\bibitem[2005]{butler} Butler, D. J., \& Martinez-Delgado D. 2005,
\aj, 129, 2217 
\bibitem[2001]{cioni} Cioni, M.-R. L., Marquette, J. B., Loup, C., et al. 2000,
\aap, 377, 945 
\bibitem[2005]{corr} Corradi, R. L. M., Magrini, L., Greimel, R., et al. 2005, 
\aap, 431, 555
\bibitem[2003]{cutri} Cutri, R. M., Skrutskie, M. F., Van Dyk, S., et al. 2003,
Expl. Supplement to the 2MASS All Sky Data Release, 
http://www.ipac.caltech.edu/2mass/releases/allsky/doc/explsup.html
\bibitem[2000]{dolphin} Dolphin, A. E., 2000, 
\apj, 531, 804
\bibitem[1989]{feast} Feast, M.W., Glass, I.S., Whitelock, P.A., et al. 1989, 
\mnras, 241, 375
\bibitem[1982]{FoxW82} Fox, M.W., \& Wood, P.R. 1982, 
\apj, 259, 198
\bibitem[2005]{fraser} Fraser, O. J., Hawley, S. L., Cook, K. H., \& Keller, S. C. 2005,
\aj, 129, 768 
\bibitem[2008]{fraser2} Fraser, O. J., Hawley, S. L., Cook, \& K. H. 2008,
\aj, 136, 1242 
\bibitem[2002]{gir} Girardi, L., Bertelli, G., Bressan, et al. 2002, 
\aap, 391, 195
\bibitem[2005]{galfalk} G\r{a}lfalk, M. 2005, NOT Annual report 2004, p18
\bibitem[2010]{geha} Geha, M., van der Marel, R. P., Guhathakurta, P., et al. 2010,
\apj, 711, 361
\bibitem[1996]{groene3} Groenewegen, M. A. T. \& Whitelock P.A. 1996,
\mnras, 281, 1347
\bibitem[2004]{Groen04} Groenewegen, M. A. T. 2004, 
\aap, 425, 595
\bibitem[2005]{groene} Groenewegen, M. A. T. 2005,
ASP Conf. Ser., arXiv:astro-ph/0506381
\bibitem[2007]{groene2} Groenewegen, M. A. T. 2007,
ASP Conf. Ser., Vol. 378, p. 433
\bibitem[1997]{han} Han, M., Hoessel, J. G., Gallagher III, J. S., et al. 1997,
\aj, 113, 1001 
\bibitem[1983]{iben} Iben, Jr., I. \& Renzini, A. 1983, 
\araa , 21, 271
\bibitem[2004]{ita} Ita, Y., Tanab\'{e}, T., Matsunaga, N., et al. 2004,
\mnras, 347, 720
\bibitem[2005]{kang} Kang, A., Sohn, Y.-J., Rhee, J., et al., 2005,
\aap, 437, 61
\bibitem[2003]{kiss} Kiss, L. L. \& Bedding, T. R. 2003, 
\mnras, 343, L79
\bibitem[2002]{lzschult} Lebzelter, T., Schultheis, M., \& Melchior, A. L. 2002,
\aap, 393, 573 
\bibitem[2005]{lzwood} Lebzelter, T., \& Wood, P. R. 2005,
\aap, 441, 1117 
\bibitem[2007]{lzwood07} Lebzelter, T., \& Wood, P. R. 2007,
\aap, 475, 643 
\bibitem[1993]{lee} Lee, M. G., Freedman, W. L., \& Madore, B. F. 1993, 
\apj, 417, 553
\bibitem[1998]{MDA} Mart\'{i}nez-Delgado, D. \& Aparicio A., 1998, 
\aj, 115, 1462
\bibitem[1998]{MDA2} Mart\'{i}nez-Delgado, D., Aparicio A. \& Gallart, C., 1999, 
\aj, 118, 2229
\bibitem[1998]{mateo} Mateo, M. 1998,
\araa, 36, 435 
\bibitem[2003]{mou} Mouhcine, M. \& Lan\c{c}on, A. 2003, 
\mnras, 338, 572
\bibitem[2009]{nich2} Nicholls, C. P., Wood, P. R., Cioni M.-R. L., et al. 2009,
\mnras, 399, 2063
\bibitem[2010]{nich} Nicholls, C. P., Wood, P. R., \& Cioni M.-R. L. 2010,
\mnras, 405, 1770 
\bibitem[2003]{wn} Nowotny, W., Kerschbaum, F., Olofsson, H., et al., 2003,
\aap, 403, 93 (Paper\,I)
\bibitem[2010]{nowo2} Nowotny, W., H\"{o}fner, S., \& Aringer, B., 2010,
\aap, 514, A35
\bibitem[2009]{pietr} Pietrzy\'{n}sky, G., Thompson, I. B., Graczyk, D., et al., 2009,
\aj, 697, 862
\bibitem[2007]{piet} Reegen, P. 2007,
\aap, 467, 1353 
\bibitem[2003]{rejka} Rejkuba, M., Minniti, D., \& Silva, D. R., 2003, 
\aap, 406, 75 
\bibitem[2004]{rejk} Rejkuba, M. 2004,
\aap, 413, 903
\bibitem[2010]{riebel} Riebel, D., Meixner, M., Fraser, O., et al., 2010,
\apj, 723, 1195
\bibitem[1994]{sage} Sage, L. J., Welch, G. A., \& Mitchell, G. F. 1998, 
\aj, 507, 726
\bibitem[2003]{schuller} Schuller, P., Ganesh, S., Messineo, M., et al., 2003,
\aap, 403, 955
\bibitem[2006]{sohn} Sohn, Y.-J., Kang, A., Rhee, J., et al., 2006,
\aap, 445, 69
\bibitem[2004a]{soszy_a} Soszy\'{n}ski, I., Udalski, A., Kubiak, M., et al. 2004a,
\actaa, 54, 129
\bibitem[2004b]{soszy_b} Soszy\'{n}ski, I., Udalski, A., Kubiak, M., et al. 2004b,
\actaa, 54, 347
\bibitem[1988]{stetson} Stetson, P. B., \& Harris, W. E. 1988,
\aj, 96, 909 
\bibitem[1998]{vanbergh} van den Bergh S. 1998, 
\aj, 116, 1688
\bibitem[1990]{wooda} Wood, P. R. 1990,
in From Miras to Planetary Nebulae, ed. M. O. Mennessier, \& A. Omont 
(Gif-sur-Yvette: \'{E}ditions Fronti\`{e}res), p 67 
\bibitem[1999]{woodb} Wood, P. R. 1999,
IAU Symp. 191, 151
\bibitem[2000]{woodc} Wood, P. R. 2000,
\pasa, 17, 18
\bibitem[2009]{wooni} Wood, P. R., \& Nicholls, C. P. 2009,
\apj, 707, 573
\bibitem[1997]{yolo} Young, L. M., \& Lo, K. 1997, 
\aj, 476, 127 
\end{thebibliography}
\end{document}